in the version submitted earlier. It can be found by typing 

\documentstyle[12pt,cite]{article}

\newcommand{\kslash}{{\rm k}\hspace{-0.53 em}\raisebox{0.15 ex}{/}}
\newcommand{\qslash}{{\rm q}\hspace{-0.51 em}\raisebox{-0.23 ex}{/}}

\title
{Electron--Electron Bound States in QED$_3$\thanks{Supported in part by
Conselho Nacional de Desenvolvimento Cient\'{\i}fico e Tecnol\'ogico (CNPq) and
by Funda\c{c}\~ao de Amparo \`a Pesquisa do Estado de S\~ao Paulo (FAPESP),
Brazil.}}

\author
{H. O. Girotti\thanks{On leave of absence from Instituto de F\'{\i}sica,
Universidade Federal do Rio Grande do Sul, Caixa Postal 15051, 91500 - Porto
Alegre, RS, Brazil.}, M. Gomes, J. L. deLyra,\\ R. S. Mendes, J. R. S.
Nascimento\\ Instituto de F\'{\i}sica, Universidade de S\~ao Paulo\\ Caixa
Postal 20516, 01498 - S\~ao Paulo, SP, Brazil.\\ and\\ A. J. da Silva\thanks{On
leave of absence from Instituto de F\'{\i}sica, Universidade de S\~ao Paulo,
Caixa Postal 20516, 01498 - S\~ao Paulo, SP, Brazil.}\\ Center for Theoretical
Physics,\\ Massachusetts Institute of Technology\\ Cambridge, MA 02139.}

\date{September 1992}

\begin{document}\maketitle

\begin{abstract}

This paper is dedicated to the study of the existence and the properties of
electron-electron bound states in QED$_3$. A detailed analysis of the infrared
structure of the perturbative series of the theory is presented. We start by
analyzing the two-point Green's function, in the Bloch-Nordsieck approximation.
The theory appears to be plagued by severe infrared divergencies, which
nevertheless disappear when vacuum-polarization effects are non-perturbatively
taken into account. The dynamical induction of a Chern-Simons term is at the
root of this mechanism.

{}From the inspection of the electron-electron non-relativistic potential it
then
follows that equally charged fermions may either repel or attract and,
moreover, that bound states do in fact exist in the theory. We calculate
numerically the binding energies and average radius of the bound states. We
find an accidental quasi-degeneracy of the ground state of the system, between
the lowest-energy $l=-3$ and $l=-5$ states, which could be related to a
radio-frequency resonance in high-$T_c$ superconductors.

\end{abstract}

\newpage

\section{Introduction}\label{sec1}

As is well-known \cite{Ja}, the Maxwell-Chern-Simons (MCS) theory is a
$(2+1)$-dimensional model describing the coupling of charged fermions
$(\bar{\psi},\psi)$ of mass $m$ and electric charge $e$ to the electromagnetic
potential $A_{\mu}$ via the Lagrangian density

\begin{eqnarray}\label{eqn1}
{\cal L}&=&-\frac{1}{4}F_{\mu\nu}F^{\mu\nu}+\frac{\theta}{4}
\epsilon^{\mu\nu\alpha}F_{\mu\nu}A_{\alpha}-\frac{1}
{2\lambda}(\partial_{\mu}A^{\mu})(\partial_{\nu}A^{\nu})\nonumber\\
&&+\frac{i}{2}\bar{\psi}\gamma^{\mu}\partial_{\mu}\psi-
\frac{i}{2}(\partial_{\mu}\bar{\psi})\gamma^{\mu}\psi-m\bar{\psi}\psi
+e\bar{\psi}\gamma^{\mu}A_{\mu}\psi ,
\end{eqnarray}

\noindent
where $F_{\mu\nu}\equiv\partial_{\mu}A_{\nu}-\partial_{\nu}A_{\mu}$, $\theta$
is a topological parameter with dimension of mass and $\lambda$ is a
gauge-fixing parameter. Neither parity nor time reversal are, separately,
symmetries of the model\footnote{Throughout this paper we use natural units
$(c=\hbar=1)$. Our metric is $g_{00}=-g_{11}=-g_{22}=1$, while for the
$\gamma$-matrices we adopt the representation
$\gamma^{0}=\sigma_{3},\gamma^{1}=i\sigma^{1},\gamma^{2}=i\sigma^{2}$;
$\sigma^{i},i=1,2,3$ are the Pauli spin matrices.}. The lowest order
perturbative contribution to the effective fermion-fermion low-energy
potential, arising from (\ref{eqn1}), was recently computed \cite{Gi} and reads

\begin{equation}\label{eqn2}
V(\vec r)=\frac{e}{2\pi}\left(1-\frac{\theta}{m}\right)K_{0}
\left(|\theta|r\right)-\frac{e}{\pi\theta}\frac{1}{mr^{2}}
\left[1-|\theta|r K_{1}\left(|\theta|r\right)\right]L.
\end{equation}

\noindent
Here, $\vec r$ is the relative distance between the electrons, $r=|\vec r|$,
$L=r_1p_2-r_2p_1$ is the orbital angular momentum, whose eigenvalues are
denoted by $l$, $\vec p$ is the relative linear momentum of the electrons, and
$K_0$ and $K_1$ designate the modified Bessel functions\footnote{The potential
(\ref{eqn2}) looks similar to the one derived by Kogan \cite{K} for the same
problem. There are some important differences, however, that we would like to
stress. The derivation in Ref.~\cite{Gi} is solely based on relativistic
quantum field theory \cite{Sk}. On the other hand, in Ref.~\cite{K} the
corresponding potential is determined in two steps. First, the $A_\mu$ vector
created by a point charge is computed. Then, the quantum mechanical Hamiltonian
describing the low energy relative motion of the two electrons is assumed to be
that of a charged particle in the presence of the external field $A_\mu$. As a
consequence, an extra term proportional to
$\left[1-|\theta|rK_{1}\left(|\theta|r\right)\right]^{2}$ arises in the
formulation of Ref.~\cite{K}.}.

The linear dependence of $V$ on $L$ accounts for the breaking of parity and
time reversal invariance in the non-relativistic approximation. The aim in
Ref.~\cite{Gi} was to determine whether the potential (\ref{eqn2}) could bind a
pair of identical fermions. For positive values of $\theta$, a numerical
solution of the Schr\"odinger equation confirmed the existence of a bound state
for $e^2/(\pi\theta)=500$, $m/\theta=10^5$ and $l=1$ . Further numerical
analysis indicated that all identical-fermion bound states are located in the
region $e^2/(\pi\theta)>1$.

Thus, we were naturally led to consider the limit $\theta\rightarrow 0$ where
the MCS theory degenerates into QED$_3$, whose Lagrangian is

\begin{equation}\label{eqn3}
{\cal L}=-\frac{1}{4}F_{\mu\nu}F^{\mu\nu}-\frac{1}{2\lambda}
(\partial_{\mu}A^{\mu})(\partial_{\nu}A^{\nu})+\frac{i}{2}
\bar{\psi}\gamma^{\mu}\partial_{\mu}\psi
-\frac{i}{2}(\partial_{\mu}\bar{\psi})\gamma^{\mu}\psi-m\bar{\psi}\psi
+e\bar{\psi}\gamma^{\mu}A_{\mu}\psi.
\end{equation}

\noindent
Power counting indicates that the S-matrix elements of QED$_3$ are plagued with
infrared singularities whose degree of divergence grows with the order of the
perturbative expansion. The situation is, therefore, more severe than in
QED$_4$ where all (on-shell) infrared divergences are logarithmic. Hence, the
usual mechanism (soft bremsstrahlung) for cancelling infrared divergences at
the level of the cross section fails, since it would only be operative for the
leading divergence. Furthermore, a non-perturbative analysis of the two-point
fermion Green function $G(p)$, carried out within the Bloch-Nordsieck (BN)
[\citen{Bl,Bo,Sz,So}] approximation, shows that $G$ is well defined for generic
values of the momentum $p$ but develops an essential on-shell singularity.
Through a multiplicative renormalization, $G\rightarrow G_R=Z^{-1}G$, one may
define a new Green function behaving on shell as a simple pole. However, $G_R$
is found to exhibit an essential off-shell singularity which reintroduces the
difficulty for the scattering amplitudes. All these developments are presented
in Sec.~\ref{sec2}.

We recall next that the BN approximation does not take into account vacuum
polarization effects. This does not constitute a serious limitation for
QED$_4$, as far as the infrared structure of the theory is concerned. In fact,
for QED$_4$ these diagrams are, on the one hand, free from infrared
singularities while, on the other hand, they do not give rise to a mechanism
for dynamical mass generation. In contrast to this, when vacuum polarization
effects are non-perturbatively incorporated into the photon propagator, the
infrared behavior of QED$_3$ changes drastically, as is demonstrated in
Sec.~\ref{sec3}. Essentially, the theory is reformulated in terms of a massive
vector boson whose mass is $|\theta |=e^2/(8\pi)$, the dynamically induced
Chern-Simons term \cite{Ja,Re} being at the root of this mechanism. Thus, the
infrared disease afflicting perturbative QED$_3$ is cured and one is allowed to
read off the corresponding effective electron-electron low-energy potential
directly from Eq.~(\ref{eqn2}), after replacing $\theta\rightarrow-e^2/(8\pi)$.
Hence,

\begin{equation}\label{eqn4}
eV^{\rm QED_{3}}(\vec r)=\frac{e^2}{2\pi}\left(
1+\frac{e^2}{8\pi m}\right)K_{0}\left(\frac{e^2}{8\pi}r\right)
+\frac{8}{mr^{2}}\left[1-\frac{e^2}{8\pi}r K_{1}\left(\frac{e^2}{8\pi}
r\right)\right]L.
\end{equation}

The terms proportional to $K_{0}$ in Eq.~(\ref{eqn4}) are both repulsive. The
term $8L/(mr^2)$ becomes attractive (repulsive) for negative (positive)
eigenvalues of $L$, while the term proportional to $K_{1}$ acts in the opposite
way. Hence, unlike the case of QED$_4$, electrons may repel or attract in
QED$_3$, the infrared structure of the theory being responsible for this
unusual behavior of the interaction between equally charged particles. A
numerical investigation on whether the potential (\ref{eqn4}) can sustain
electron-electron bound states is presented in Sec.~\ref{sec4}. There we show
that such bound states do in fact exist. For some of them the corresponding
dissociation temperature will be seen to be in qualitative agreement with the
values encountered for the critical temperature of high-$T_c$ superconductors.

Our conclusions are contained in Sec.~\ref{sec5}. A preliminary version of
some of the results presented in this paper is contained in Ref.~\cite{p}.

We remind the reader that the signs of the electron's spin and of the induced
topological mass are both equal to the sign of the fermionic mass term in the
Lagrangian (\ref{eqn3}) \cite{Ja}. Thus, in a theory involving both spin-up and
spin-down fermions there will be no induced Chern-Simons term and, for the
specific case of electrodynamics, there will be no cure for the infrared
problem. This explains why it is essential that we restrict ourselves, in this
paper, to the case of a single two-component fermion field. The generalization
to the case of several flavors is straightforward and will be mentioned along
the way.

\section{Infrared Structure}\label{sec2}

As we already mentioned in the Introduction, power counting alone reveals that
perturbative QED$_3$ is afflicted by severe on-shell infrared singularities. To
gain some insight into the infrared structure of this theory, we shall start by
computing the two-point fermionic Green function in the BN approximation. To
keep the singularities under control, we approach QED$_3$ as the
$\theta\rightarrow 0$ limit of the MCS theory (\ref{eqn1}). Clearly, both
theories exist off shell and MCS goes into QED$_3$ as $\theta\rightarrow 0$.

The BN approximation consists of the replacement of the $\gamma^\mu$-matrices
in Eq.~(\ref{eqn1}) by a vector $u^\mu$, with $u^2=1$. As a consequence, the
usual free fermion propagator is replaced by the retarded function

\begin{eqnarray}\label{eqn6}
G_F(x-y)&\equiv&\frac{1}{(2\pi)^3}\int d^3p\,\frac{{\rm
e}^{-ip\cdot(x-y)}}{u\cdot p-m+i\epsilon}\nonumber\\&=&-i u^{\scriptstyle
0}H(x^{\scriptstyle 0}-y^{\scriptstyle 0}){\rm e}^{-i\frac{m}{u^{\scriptstyle
0}}(x^{\scriptstyle 0}-y^{\scriptstyle 0})}
\delta^{(2)}\left[
\vec u(x^{\scriptstyle 0}-y^{\scriptstyle 0})-u^{\scriptstyle 0}(\vec x
-\vec y)\right],
\end{eqnarray}

\noindent
where $H(x^{\scriptstyle 0})$ is the Heaviside step function. Then, the
interaction does not imply any corrections to the vector meson propagator. For
our present purposes this does not represent a serious drawback since vacuum
polarization insertions do not alter the leading infrared behavior of a graph.

The Green functions of the theory will be computed by functionally
differentiating the generating functional $U_{\scriptstyle
0}[J_\mu,\,\eta,\bar\eta]$ with respect to the external sources. In this paper,
$J_\mu$ and $\eta$, $\bar\eta$ denote the vector meson and fermion external
sources, respectively. After integration on the fermionic degrees of freedom
one finds

\begin{eqnarray}\label{eqn7}
U_{0}[J_\mu,\eta,\bar\eta]&=&{\cal N}\int[{\cal D}A_\mu]
{\rm e}^D\exp\left\{iS[A]-\frac{i}{2\lambda}
\int d^3x(\partial_\mu A^\mu)^2\right.\nonumber\\
&&+\left.i\int d^3x\,J_\mu(x)A^\mu(x)-
i\int d^3x\int d^3y\,\bar\eta(x)G[A\vert x,y]\eta(y)\right\},
\end{eqnarray}

\noindent
where ${\rm e}^D$ is the fermionic determinant, $S[A]=-F_{\mu\nu}F^{\mu\nu}/4$
and $G[A\vert x,y]$ is the fermionic two-point Green function in the presence
of an external field $A_\mu$.

The solubility of the model in the BN approximation is partly due to the fact
that the $A_\mu$ propagator is not corrected by the interaction, and partly due
to the factorization property of $G[A\vert x,\,y]$. In fact, one can convince
oneself that the differential equation

\[
\left\{iu^\mu\left[{\partial_{x}}_{\mu}-ieA_\mu(x)\right]-m\right\}G[A\vert
x,\,y]=\delta^{(3)}(x-y)
\]

\noindent
is solved by

\begin{equation}\label{eqn8}
G[A\vert x,\,y]=h[A\vert x]G_F(x-y)h^{-1}[A\vert y],
\end{equation}

\noindent
where

\[
h[A\vert x]=\exp\left[ie\int d^3z\,\xi^\mu(x-z)A_\mu(z)\right],
\]

\noindent
and

\[
\xi^\mu(x)=\frac{i}{(2\pi)^3}\int d^3k\frac{u^\mu}{(u\cdot k)}
{\rm e}^{-ik\cdot x}.
\]

\noindent
One can check that expression (\ref{eqn8}) can be cast in a form analogous to
that quoted in Ref.~\cite{Bo} for the case of QED$_4$.

Formally, the computation of $D$ yields $D=\int_{0}^{e}de^{\prime}\int
d^3xG[A\vert x,x]u^\mu A_\mu(x)$, which in view of (\ref{eqn8}) reduces to

\[
D=e\int d^3x\,G_F(x,x)u^\mu A_\mu(x),
\]

\noindent
showing that $D$ can, at most, depend linearly on $A_\mu$. However, as seen
from (\ref{eqn6}), $G_F(x,x)$ is ambiguous. In the original model, with
$\gamma^\mu$ instead of $u^\mu$, Lorentz invariance demands the vanishing of
the tadpole contribution to the fermionic determinant. Based on this fact we
shall therefore take, from now on, $D=0$.

The complete two-point fermion Green function can now be readily found,

\begin{eqnarray*}
G(x,y;m,\theta)&\equiv&\left.\frac{1}{U_{0}[0]}\,\,
\frac{1}{i^{2}}\frac{\delta^{2}U_{0}[J_{\mu}=0,\eta,\bar\eta]}
{\delta\bar\eta(x)\,\delta\eta(y)}\right\lfloor_{\eta=\bar\eta=0}\nonumber\\
&=&iG_{F}(x-y)\exp\left[-\frac{i}{2}e^{2}\int d^3z\int d^3z^{\prime}
S_{\mu}(x,y;z)\Delta^{\mu\nu}(z,z^{\prime}) S_{\nu}(x,y;z^{\prime})\right],
\end{eqnarray*}

\noindent
where

\[
S^{\mu}(x,y;z)\equiv\xi^\mu(x-z)-\xi^\mu(y-z),
\]

\noindent
and $i\Delta^{\mu\nu}$ is the photon propagator. From (\ref{eqn7}) it follows
that, in momentum space,

\begin{equation}\label{eqn9}
i\tilde\Delta^{\mu\nu}(k,\theta)=-\frac{i}{k^2-\theta^{2}}\left(
g^{\mu\nu}-\frac{k^{\mu}k^{\nu}}{k^2}\right)
-\frac{\theta}{k^2-\theta^{2}}\frac{\epsilon^{\mu\nu
\rho}k_{\rho}}{k^2}-i\lambda\frac{k^{\mu}k^\nu}{(k^2)^2}.
\end{equation}

\noindent
Hence, for the Fourier transform $G(p;m,\theta)\equiv\int
d^3x\exp(ipx)G(x;m,\theta)$ one gets, after some algebra,

\[
G(p;m,\theta)=\int_{0}^{\infty}d\nu\exp\left[i\nu(u\cdot p-m+i\epsilon)+
f(\nu,\theta)\right],
\]

\noindent
where

\begin{equation}\label{eqn11}
f(\nu,\theta)=
-\frac{ie^2}{(2\pi)^3}\int d^3k\{1-\cos[\nu(u\cdot k)]\}U(k,\theta),
\end{equation}

\noindent
and

\[
U(k,\theta)\equiv
\frac{u^{\mu}\tilde\Delta_{\mu\nu}(k,\theta)u^{\nu}}{(u\cdot k)^2}.
\]

\noindent
Since $u^\mu\epsilon_{\mu\nu\rho}u^\nu=0$, the second term in (\ref{eqn9}) does
not contribute to $U(k,\theta)$. Thus, the $\theta\rightarrow 0$ limit of $G$
in the MCS theory is the same as that of the corresponding function in massive
QED$_3$.

The behavior of the integrand in Eq.~(\ref{eqn11}) in the limit
$k\rightarrow\infty$, with $u\cdot k$ kept constant, tells us that
$f(\nu,\theta)$ develops a logarithmic ultraviolet divergence. This divergence,
which can be absorbed into a mass renormalization, is entirely due to the
approximation being used ($u^{\mu}$ instead of $\gamma^{\mu}$)\footnote{A
similar situation arises in QED$_4$ where the corresponding function $f(\nu)$
presents linear and logarithmic ultraviolet divergences, the linear divergence
being induced by the replacement $\gamma^{\mu}\rightarrow u^{\mu}$ \cite{Bo}.}.
The two-point Green function $G$ in terms of the renormalized mass $m_R$ turns
out, then, to be given by

\begin{eqnarray}\label{eqn13}
\lefteqn{G(p; m_{R},\theta)=\frac{1}{\theta}\int_{0}^{\infty}dv}
&&\nonumber\\&&\!\!\!\!\!\!
\exp\left\{iv\left[i\epsilon+\frac{Q}{\theta}-\frac{e^2}{4\pi\theta}
\left(\ln\frac{M}{\theta}+\frac{{\rm e}^{-iv}-1}{iv}+
2\frac{{\rm e}^{-iv}-1+iv}{(iv)^2}+Ei(-iv)-\lambda\right)\right]\right\},
\end{eqnarray}

\noindent
where $Q\equiv u\cdot p-m_R$, $Ei(z)$ is the exponential-integral function,

\begin{equation}\label{eqn14}
m_R=m+\frac{e^2}{4\pi}\ln\left (\frac{\Lambda}{M}\right ),
\end{equation}

\noindent
$\Lambda$ is an ultraviolet cutoff, and $M$ plays the role of a subtraction
point. For $\theta\not=0$, the singularity structure of $G(p;m_R,\theta)$ can
be seen to consist of a pole at\footnote{An analogous situation arises in the
one-loop calculation of Ref.~\cite{Ja}, where a gauge-dependent contribution
to the electron mass is found.}

\[
u\cdot p=m_R+\frac{e^2}{4\pi }\left(\ln\frac{M}{\theta}-\lambda \right),
\]

\noindent
plus branch cuts. If one insists in keeping the pole of $G$ at the renormalized
mass defined in Eq.~(\ref{eqn14}), one must choose $M=\theta\exp(\lambda)$,
thus
lumping all the gauge dependence of $G$ into the subtraction point. Moreover,
the Green function $G$ can be renormalized,

\[
G(p;m_R,\theta)\rightarrow G_R(p;m_R,\theta)\equiv Z^{-1}G(p;m_R,\theta),
\]

\noindent
such that

\[
\lim_{u\cdot p\to m_R}(u\cdot p-m_R)G_R(p; m_R,\theta)=i.
\]

\noindent
It may be checked that the ultraviolet-finite renormalization constant

\[
Z=\exp\left(-\frac{e^2}{4\pi \theta}\right)
\]

\noindent
does the job. Everything runs smoothly for $\theta\not=0$.

However, the above scheme does not survive the limit $\theta\rightarrow 0$.
While the pole of $G_R$ can be kept at the value of the renormalized mass, the
constant $Z$ presents an essential singularity which in turn is transferred to
$G_R$ and shows up for all off-shell values ($u\cdot p\not=m_R$) of the
momentum. One may try to circumvent this difficulty by defining the two-point
fermion Green function of QED$_3$ as the $\theta\rightarrow 0$ limit of
(\ref{eqn13}), keeping $M$ arbitrary. For $Q<0$ and after a Wick-like rotation
one arrives at

\begin{equation}\label{eqn18}
G(p;m_R,\theta=0)=\frac{i}{Q}\int_{0}^{\infty}dv\exp\left\{-v+\frac{e^2}
{4\pi Q}v\left[C+\ln\left(-\frac{M}{Q}v\right )-\lambda\right]\right\},
\end{equation}

\noindent
where $C$ is the Euler constant. On the other hand, if $Q>0$ one obtains

\begin{equation}\label{eqn19}
G(p;m_R,\theta=0)=-\frac{i}{Q}\int_{0}^{\infty}dv\exp\left\{v-\frac{e^2}
{4\pi Q}v\left[C+\ln\left(\frac{M}{Q}v\right)-\lambda\right]\right\},
\end{equation}

\noindent
although in this last case the Wick-like rotation is only allowed if
$e^2\not=0$. Expressions (\ref{eqn18}) and (\ref{eqn19}) are well defined for
generic values of $Q$ but contain an essential singularity at $Q=0$\footnote{It
is instructive to compare this result with the corresponding one in QED$_4$,
where the two-point fermion Green function exhibits a power law behavior in the
variable $Q$ \cite{Bo}.}.

In both cases above, the BN approximation fails to provide an acceptable
functional form for the two-point fermion Green function of QED$_3$. This is
an indication that the infrared singularities occurring in each order of
perturbation theory do not add up to a finite limit. Moreover, the inclusion of
soft bremsstrahlung emission, which in QED$_4$ removes the infrared problem
from the cross sections, is of no help in the present case since it would only
act on the leading divergences. Nothing changes, qualitatively, if vacuum
polarization insertions are added order by order in perturbation theory.
Nevertheless, when vacuum polarization effects are non-perturbatively
incorporated into the photon propagator, a photon mass is dynamically generated
and as a consequence the infrared disease is cured. To show how this comes
about is the purpose of the next section.

\section{Vacuum Polarization Effects}\label{sec3}

The lowest-order graph contributing to the vacuum polarization tensor
$\Pi^{\rho\sigma}$ yields

\[
\Pi^{\rho\sigma}(q)=ie^2\int\frac{d^3k}{(2\pi)^3}\frac{{\rm
Tr}[\gamma^\rho(\kslash+\qslash+m)\gamma^\sigma(\kslash
+m)]}{[(q+k)^2-m^2](k^2-m^2)}.
\]

\noindent
Since we are interested in the quantum corrections to a non-relativistic
potential, we shall retain only those terms which are zero and first order in
$q$. Gauge invariance alone ensures that $\Pi^{\rho\sigma}(0)=0$. For the
first-order contribution, which gives origin to the induced Chern-Simons term
\cite{Ja,Re}, one finds

\[
\Pi_{\rho\sigma}^{(1)}(q)=-i\frac{e^2}{8\pi}\epsilon_{\rho\sigma\mu}q^{\mu}.
\]

\noindent
We emphasize that $\Pi_{\rho\sigma}^{(1)}(q)$ is ultraviolet-finite and that,
therefore, no regularization is needed for its computation. Alternatively, if
one adopts the point of view that ultraviolet divergences should be kept under
control by the regularization of the entire theory, the result quoted in this
last equation is true only if a parity-time reversal invariant regularization
is used.

We now modify the photon propagator by resumming the geometric series resulting
from the iteration of $\Pi_{\rho\sigma}^{(1)}$ (see Fig.~\ref{fig01}). This is
formally equivalent to adding to the Lagrangian (\ref{eqn3}) the term

\[
{\cal L}_{CS}=-\frac{e^2}{32\pi}\epsilon^{\mu\nu\rho}F_{\mu\nu}
A_{\rho},
\]

\noindent
which is nothing but a Chern-Simons term with and induced topological mass

\begin{equation}\label{eqn22}
\theta_{in}=-\frac{e^2}{8\pi}.
\end{equation}

\noindent
Hence, in the low-energy regime, the theory effectively describes the
interaction of fermions with massive vector particles of mass
$\vert\theta_{in}\vert$ and is, therefore, free from infrared problems.
Moreover, from Eqs.~(\ref{eqn2}) and (\ref{eqn22}) it follows that the QED$_3$
effective electron-electron low-energy potential is, in fact, the one given in
Eq.~(\ref{eqn4}).

We verify the consistency of the present formulation of QED$_3$ by showing that
all remaining contributions to $V^{\rm QED_{3}}$ are, up to some power of
$\ln(e^2/m)$, of order $e^2/m$ or higher with respect to (\ref{eqn4}), and
vanish as $e^2/m\rightarrow 0$. To see how this comes about, we compute the
vertex correction $\Lambda^\mu$ to $V^{\rm QED_{3}}$ arising from the diagram
in which only one massive vector particle is exchanged\footnote{For the vertex
corrections in MCS theory see Ref.~\cite{Ko}. We would like to point out,
however, that the limit $\theta\rightarrow 0$ in the MCS theory differs
radically from the limit $\theta_{in}=-e^2/(8\pi)\rightarrow 0$ in QED$_3$,
since in the latter case the vanishing of the topological mass also implies the
vanishing of the coupling constant.} (see Fig.~\ref{fig02}). After the
replacement $\theta\rightarrow\theta_{in}$, the massive vector field propagator
can be read off directly from Eq.~(4) of Ref.~\cite{Gi},

\[
D_{\mu\nu}(k)=\frac{-i}{k^{2}-\theta^{2}_{in}}\left
(P_{\mu\nu}\,\,-\,\,i\theta_{in}\epsilon_{\mu\nu\rho}\frac
{k^{\rho}}{k^{2}}\right)-i\lambda\frac{k_{\mu}k_{\nu}}{k^{4}}
f(k^{2}),
\]

\noindent
where $P_{\mu\nu}\equiv g_{\mu\nu}-{k_{\mu}k_{\nu}}/{k^{2}}$, and an arbitrary
function $f(k^{2})$ has been incorporated into the longitudinal part.
Accordingly, $\Lambda^\mu$ can be split as follows

\[
\Lambda^\mu=\Lambda^{\mu}_{g}+\Lambda^{\mu}_{\epsilon}+
\Lambda^{\mu}_{L},
\]

\noindent
where the subscripts $g$, $\epsilon$, and $L$ make reference to those
pieces of $D_{\mu\nu}$ proportional to $g_{\mu\nu}$,
$\epsilon_{\mu\nu\rho}k^{\rho}$ and $k_{\mu}k_{\nu}$, respectively. The
computation of $\Lambda^{\mu}_{L}$ is straightforward and yields

\[
\Lambda^{\mu}_{L}=i\lambda\bar v^{(+)}({\bf p}_{1}^{\prime})\gamma^{\mu}
v^{(-)}({\bf p}_{1})\int\frac{d^{3}k}{(2\pi^{3})}\frac{f(k^{2})}{k^2},
\]

\noindent
where, as in Ref.~\cite{Gi}, $v^{(-)}({\bf p}_{1})$ ($\bar v^{(+)}({\bf
p}_{1}^{\prime})$) is a two-component spinor describing a free electron of
two-momentum ${\bf p}_{1}$ (${\bf p}_{1}^{\prime}$) in the initial (final)
state. Since $\Lambda^{\mu}_{L}$ does not depend on the momentum transfer
$q\equiv p_{1}^{\prime}-p_{1}$, it can be absorbed into the wave-function
renormalization constant. When computing $\Lambda^{\mu}_{g}$ and
$\Lambda^{\mu}_{\epsilon}$ only zero- and first-order terms in ${q}/{m}$ will
be retained, since our interest here is restricted to the non-relativistic
regime. Furthermore, all the momentum integrals over loops are
ultraviolet-finite and there is no need for regularization. After absorbing the
zero-order terms into the wave-function renormalization constant, one arrives
at

\begin{equation}\label{eqn23}
\Lambda^{\mu (1)}_{g}(q)=\frac{N_{g}}{16\pi}
\frac{e^2}{m}\epsilon^{\mu\nu\rho}
\frac{q_{\nu}}{m}\,\,\bar v^{(+)}({\bf p}_{1}^{\prime})
\gamma_{\rho}v^{(-)}({\bf p}_{1}),
\end{equation}

\noindent
and

\begin{equation}\label{eqn24}
\Lambda^{\mu(1)}_{\epsilon}(q)=
-\frac{N_{\epsilon}}{16\pi}\frac{e^2}{m}
\epsilon^{\mu\nu\rho}\frac{q_{\nu}}{m}\,\,\bar
v^{(+)}({\bf p}_{1}^{\prime})\gamma_{\rho}v^{(-)}({\bf p}_{1}),
\end{equation}

\noindent
where

\begin{equation}\label{eqn25}
N_{g}=-3+3\frac{e^2}{8\pi m}+2\left[1-
\frac{3e^{4}}{(16\pi m)^{2}}\right]\ln\left(1+\frac{16\pi m}{e^2}\right),
\end{equation}

\noindent
and

\begin{equation}\label{eqn26}
N_{\epsilon}=-\frac{e^2}{\pi m}\ln\left(1+\frac{16\pi m}{e^2}\right).
\end{equation}

\noindent
{}From Eqs.~(\ref{eqn23}--\ref{eqn26}) it follows that $\Lambda^{\mu(1)}(q)$
behaves as $(e^2/m)\ln(e^2/m)$ when $e^2/m\rightarrow 0$
(${\theta_{in}}/{m}\rightarrow 0$), while power-counting indicates that the
full vertex insertion $\Lambda^{\mu}(q)$ (Fig.~\ref{fig02}) may diverge
logarithmically at the just mentioned infrared limit. The fact that
$\Lambda^{\mu(1)}(q)$ exhibits an improved infrared behavior is not a
peculiarity of the particular insertion under analysis but applies to any
vertex part involving an arbitrary number of exchanged massive vector
particles. Indeed, the leading infrared divergence of any of these parts only
shows up in those terms containing even powers of the momentum transfer $q$, as
can be seen by setting to zero the loop momenta in the numerators of the
corresponding Feynman integrals. Thus, the terms linear in $q$ exhibit a milder
infrared behavior.

By using the technique described in Ref.~\cite{Gi} one finds the correction
$\Delta V^{\rm QED_3}$ arising from the diagrams in Fig.~\ref{fig03},

\begin{eqnarray}\label{eqn27}
e\Delta V^{\rm QED_3}&=&
\frac{e^2}{8\pi}\frac{e^2}{m}(N_g-N_{\epsilon})\nonumber\\
&\times&\left[\frac{e^2}{16\pi^2m}\left (1+\frac{e^2}{8\pi
m}\right)K_0\left(\frac{e^2}{8\pi}r\right)\right.
-\left.\frac{\delta^2({\bf r})}{m^2}+\frac{e^2}{8\pi^2m^2}K_1
\left(\frac{e^2}{8\pi}r\right)\frac{L}{r}\right].
\end{eqnarray}

\noindent
{}From Eqs.~(\ref{eqn25}--\ref{eqn27}) it follows that $e\Delta V^{\rm QED_3}$
also behaves as $(e^2/m)\ln(e^2/m)$ when $e^2/m\rightarrow 0$, and turns out to
be negligible if $e^2/m\ll 1$. This establishes the region of validity of our
approximations.

\section{Electron-Electron Bound States}\label{sec4}

We turn next to the investigation of the existence and properties of bound
states of two identical fermions of mass $m$ and charge $e$ interacting through
the non-relativistic potential $V^{\rm QED_3}$ given in (\ref{eqn4}). The
corresponding radial Schr\"odinger equation is found to read

\[
{\cal H}_{l}R_{nl}(y)=\epsilon_{nl}R_{nl}(y),
\]

\noindent
where $R_{nl}$ is the radial part of the wave function, $l$ is, as defined
before, the angular momentum eigenvalue, $n$ is the principal quantum number,
and ${\cal H}_{l}$ is the effective radial Hamiltonian, namely,

\[
{\cal H}_{l}R_{nl}(y)=-\left[\frac{\partial^{2}
R_{nl}(y)}{\partial y^{2}}+\frac{1}{y}\frac{\partial R_{nl}(y)}{\partial y}
\right]+U^{eff}_{l}\!(y)R_{nl}(y).
\]

\noindent
Here $U_{l}^{eff}$ denotes the effective radial potential arising from
(\ref{eqn4}),

\begin{equation}\label{eqn29}
U^{eff}_{l}\!(y)=\frac{l^{2}}{y^{2}}+\frac
{\alpha_{in}}{2}\left(1+\beta_{in}\right)K_{0}(y)
+\frac{\alpha_{in}l}{y^2}\left[1-y K_{1}(y)\right],
\end{equation}

\noindent
where we introduced the dimensionless parameters $y=e^2r/(8\pi)$,
$\alpha_{in}=e^2/(\pi|\theta_{in}|)=8$, $\beta_{in}=m/|\theta_{in}|=8\pi
m/e^2$, and $\epsilon_{nl}={64\pi^2mE_{nl}}/{e^4}$, $E_{nl}$ being the energy
eigenvalue.

A straightforward analysis of (\ref{eqn29}) reveals that electron-electron
bound states are possible only for $l=-1,\,-3,\,-5$ and $-7$. In order to see
this, we consider the derivative

\[
\frac{\partial U_{l}^{eff}(y)}{\partial y}=
-\frac{2l(l+\alpha_{in})}{y^3}+\frac{\alpha_{in}l}{y}K_{0}(y)
-\left[\frac{\alpha_{in}}{2}\left(1+\beta_{in}\right)
-\frac{2\alpha_{in}l}{y^2}\right]K_{1}(y).
\]

\noindent
For $l$ negative and outside the interval $[-7,-1]$, this derivative is
negative definite and, therefore, $U_{l}^{eff}$ has no local minima. Inside
this interval, on the other hand, an examination of the behavior of
$U_{l}^{eff}$ reveals that for small $y$ it diverges to $+\infty$, while for
large $y$ it approaches zero from negative values. It then follows that
$U_{l}^{eff}$ has an absolute minimum, thus opening up the possibility for
bound states. The fact that Fermi statistics requires that $l$ be odd reduces
the possibilities to the values mentioned above. Had we chosen the negative
sign for $m$ we would have obtained $\theta_{in}=+e^2/(8\pi)$ and bound states,
with identical energy eigenvalues, would be possible for $l=+1,\,+3,\,+5$ and
$+7$. This analysis can be easily extended to the case of two different
flavors. If the electrons are in an antisymmetric flavor state, bound states
may emerge for $l=-2,\,-4$ and $-6$.

The straight numerical resolution of the Schr\"odinger equation is useful for
determining quickly whether bound states exist or not, but it is not so useful
for the determination of the binding energies of such bound states. The reason
is that in order to solve the equation one needs, beforehand, a definite value
for the energy parameter. Since the energy eigenvalue is not known a priori,
one must search for it, solving the equation repeatedly by starting with the
boundary condition for $R_{nl}$ at $y=0$, and propagating the wave function
forward until one realizes the boundary condition for large $y$. Besides being
a difficult process in itself, due to the very sensitive dependence of the
asymptotic behavior of the wave function on the value of the energy parameter,
it interacts in a complex way with the finiteness of the integration interval
and of the integration range.

The existence of bound states for the potential given in (\ref{eqn29}) was
examined numerically by means of a stochastic variational algorithm. We found
this method much more convenient than the resolution of the complete eigenvalue
problem, which is large and hard. The variational algorithm consists of a
direct search for the lowest eigenvalues only, and is therefore more
appropriate for our purposes. In all cases which were examined with both this
procedure and the Schr\"odinger equation, the results were the same, within the
numerical limitations of each method. With the variational algorithm we were
able to quickly identify, for a given $l$, the state of minimum energy ($n=0$),
for a variety of values of $\beta_{in}$, and to estimate roughly the binding
energies. However, for a precise determination of the binding energies, long
double-precision runs were necessary. We also measured the expectation value
($\bar{r}_{nl}$) of the radius in the state thus obtained. A typical
configuration of $R_{0l}$, and the corresponding potential, are displayed in
Fig.~\ref{fig04}.

In the variational algorithm we varied the wave function $R_{0l}$, subject to
the boundary condition $R_{0l}(0)=0$, while trying to decrease the energy of
the state, which is given by

\[
\epsilon_{0l}=\frac{\int_{0}^{\infty}{\rm d}y\;y\left\{\left[\frac{\partial
R_{0l}(y)}{\partial y}\right]^2+U_{l}^{eff}(y)R_{0l}(y)^{2}\right\}}
{\int_{0}^{\infty}{\rm d}y\;y\;R_{0l}(y)^{2}}.
\]

\noindent
During the variation of the wave function, besides implementing the boundary
condition at the origin, we kept constant its maximum value. This implies that
the normalization of the wave function relaxes to some arbitrary value. The
requirement that the wave function should decrease exponentially for large
values of $y$ was realized automatically by the algorithm.

In order to be able to represent the wave function $R_{0l}$ and the expectation
value of the energy $\epsilon_{0l}$ numerically, we must chose a discrete
collection of points (sites) along the radial direction. Hence, we calculated
the ratio of sums

\[
\epsilon_{0l}=\frac{\sum_{i=0}^{i_{max}-1}\Delta y\;\frac{y_{i}+y_{i+1}}{2}
\left[\frac{\Delta R_{0l}(y)}{\Delta y}\right]_{i}^2
+\sum_{i=1}^{i_{max}}\Delta y\;y_i\;U_{l,i}^{eff}R_{0l,i}^{2}}
{\sum_{i=1}^{i_{max}}\Delta y\;y_i\;R_{0l,i}^{2}},
\]

\noindent
where $\Delta R_{0l}$ are forward differences, and the values of $\Delta y$ and
$i_{max}$ were chosen to ensure a big enough range of integration $Y=\Delta
y\,i_{max}$ and a small enough integration interval $\Delta y$.

We then varied the wave function randomly, accepting any changes that decreased
the energy. The function was varied by sweeping the sites, excluding the point
at the origin but including the point corresponding to $Y$, where it was left
free to fluctuate. At each site, the wave function was changed by adding to it
a random number in the interval $[-\varepsilon,\varepsilon)$. After a certain
number of sweeps (blocks), the position of the maximum was determined and the
function was renormalized to keep constant its maximum value.

During the sweeping procedure we monitored the {\em acceptance rate} of the
trial changes in the wave function, which depends strongly on the value chosen
for the range $\varepsilon$. In order to keep the code efficient, it is
necessary to ensure that the wave function is being changed as fast as
possible. The optimum value for $\varepsilon$ is the one which maximizes the
product of this range by the acceptance rate. This product is a good estimator
for the overall speed with which the wave function is being modified. We
included in the code a feedback mechanism which searched for this optimum value
of $\varepsilon$ by maximizing the product.

The runs were started with a large value for $\varepsilon$, which was then
continuously adjusted by the code as the run proceeded. We found that the
feedback mechanism worked very well to decrease $\varepsilon$ to appropriate
values but that sometimes it could overshoot and get hung at extremely small
values of the range. This was particularly true in the long runs needed for the
precise determination of the binding energies. This problem was overcome by a
procedure which we call {\em quenching}. We stopped the runs two or three
times, storing the wave function produced so far, and restarted them with a
large value for the range $\varepsilon$. This procedure greatly increased the
convergence speed of the code. Figs.~\ref{fig05} and \ref{fig06} display the
relaxation of the binding energy and of the average radius of the bound state,
at the final quenching cycle, in a particular case.

Another relevant numerical issue is a phenomenon related to the finite
numerical precision of the computer. During the sweeping procedure it was often
found that many changes in the wave function did not produce any changes at all
in the energy. Clearly, in these cases the change in the energy was below the
numerical precision available. When this happened we accepted the change
anyway, monitoring separately the acceptance rates due to relaxation of the
energy and to this phenomenon. The acceptance of these changes interacted with
the relaxation changes, increasing the efficiency of the code. Another source
of numerical errors was related to the sweeping procedure. In order to save
computation time, during the blocks of sweeps, the energy was not re-calculated
every time a site was upgraded, but had instead its variations added to it
along the way. This introduced numerical errors in the system and, in order to
keep the errors small, after each block of sweeps the energy was calculated
anew.

The finite-precision phenomena described above caused a small random drift in
the results, which were used to estimate the stochastic errors. The stability
of the results was tested against variations of $Y$ and $\Delta y$. Due to the
exponential decay of the wave function for large $y$, it was found that, given
a large enough $Y$, the results were essentially independent of it. They were
somewhat more sensitive to the finiteness of $\Delta y$, approaching a limit
linearly with $\Delta y$, as $\Delta y\rightarrow 0$. Also, the code became
progressively less efficient as one decreased $\Delta y$. The reason for this
is easily understood. For small $\Delta y$ any change in the wave function at a
single site tends to produce large spikes in its derivative and, then, the
derivative term of the energy causes all but very small changes to be rejected.
For small $\Delta y$ the local upgrading is inefficient to change the
long-wavelength components of the configurations.

Being prevented by the computational limitations from decreasing $\Delta y$
indefinitely, we estimated the $\Delta y=0$ limits of the results by linear
regression to zero from three finite-$\Delta y$ runs, for each set of values of
the parameters $\beta_{in}$ and $l$. These are the data presented in
Figs.~\ref{fig07}--\ref{fig11}. They were calculated assuming $m$ to be the
usual electron mass. The energy eigenvalues for $l=-5$ and $l=-7$ are very
close to those corresponding to $l=-3$ and $l=-1$, respectively, and for this
reason they have not been included in the figures. The error bars are too small
to be visible in the graphs. The stars mark the points calculated directly,
while the lines were obtained by splines interpolations. The corresponding
dissociation temperatures $T_{d,nl}$ are plotted in Fig.~\ref{fig09}.
Table~\ref{tbl1} contains approximate fits for the data.

An examination of $U^{eff}_{l}$ in Eq.~(\ref{eqn29}) shows that for large $y$
the potential changes very little between the cases $l=-3$ and $l=-5$, and
between the cases $l=-1$ and $l=-7$. The dominant term $l(l+8)/y^2$ does not
change at all; the only term that does change contains $K_{1}$, which is
exponentially damped for large $y$. It is not so clear how much the potential
changes near the well, which is the most important region as far as the bound
states are concerned. In fact, one finds but very small differences between the
energies for $l=-3$ and $l=-5$, and between those for $l=-1$ and $l=-7$. The
ground state of the system is the $l=-3$, $n=0$ state, but the $l=-5$, $n=0$
state is very close to it, forming a quasi-degenerate doublet.

Assuming that transitions between these states do occur, the transition energy
would correspond to very low resonant frequencies, in the radio spectrum. This
resonant frequency $f_R$ appears in Fig.~\ref{fig10}, as a function of
$\beta_{in}$, and in Fig.~\ref{fig11}, as a function of the dissociation
temperature. Presumably, the states with larger $n$ also form similar doublets
and there should be further transition frequencies, similar to this one,
corresponding to them.

\section{Conclusions}\label{sec5}

There are ranges of the parameters $l$ and $\beta_{in}$ where the results of
Sec.~\ref{sec4} are numerically consistent with the observed phase-transition
temperatures of high-$T_c$ superconductors. We do not think that this fact is
merely accidental, although we are not claiming that QED$_3$ appropriately
describes all features of high-$T_c$ superconductivity. Also note that if the
initial state of the system contains only low energy electrons, energy
conservation forbids the production of electron-positron pairs. No
electron-positron bound states can be formed and, as a consequence, only the
electron-electron bound state can characterize the many-body ground state.

Certainly, the calculations presented here are still incomplete. For instance,
one should recalculate the potential for the finite-temperature field theory,
which presumably will introduce corrections in the energies and temperatures.
It also remains to be shown that the theory exhibits a phase transition and
that the many-body ground state is a condensate of electron-electron pairs.
However, if this turns out to be the case, we find reasonable to think that the
dissociation temperatures found in this paper should be, at least, rough
approximations for the true phase-transition temperatures.

We believe that the accidental quasi-degeneracy uncovered in this work may be
important to test the relevance of the model for the description of high-$T_c$
superconductors. The corresponding transition frequency is directly related to
the binding energy of the ground state, and therefore to the predicted
dissociation temperature. The relation thus implied, between this frequency and
the phase-transition temperature, may provide a means of testing the
applicability of the model for the description of high-$T_c$ superconductors.
If the model considered here is more than superficially related to high-$T_c$
superconducting materials, it should be possible to detect such a transition.
It certainly would be very interesting to further pursue this matter.

\section*{Acknowledgements}

One of us (AJS) thanks Prof. R. Jackiw for many valuable criticisms, and the
Center for Theoretical Physics of MIT for the kind hospitality. The computer
work presented in this paper was realized mostly on the Unix computer systems
of the ``Departmento de F\'{\i}sica Matem\'atica'', which were acquired with
grants from Fapesp. The computer systems of the ``Instituto de F\'{\i}sica''
were also used.

\newpage

\begin{center}
FIGURES
\end{center}

\begin{enumerate}

\item\label{fig01}
Wavy lines represent free photons, while dashed lines refer to massive vector
particles.

\item\label{fig02}
The vertex insertion $\Lambda^{\mu}$.

\item\label{fig03}
Diagrams contributing to the potential $\Delta V^{\rm QED_3}$.

\item\label{fig04}
The bound-state wave function $R_{0l}$ (dashed line) and the dimensionless
potential $U^{eff}_{l}$ (continuous line) for $l=-3$, $\beta_{in}=2000$,
$\Delta y=0.25$ and $Y=200$. The vertical scale refers only to the potential.
The normalization of the wave function is arbitrary.

\item\label{fig05}
Relaxation of the dimensionless energy parameter $\epsilon_{0l}$ for $l=-3$,
$\beta_{in}=2000$, $\Delta y=0.25$ and $Y=200$, on the last quenching run, as a
function of the blocks of sweeps. The vertical scale is relative to the minimum
value $\epsilon_{0-3}=-0.03266121075526209$, which corresponds to the zero of
the scale.

\item\label{fig06}
Relaxation of the dimensionless average radius $\bar{y}_{0l}$ for $l=-3$,
$\beta_{in}=2000$, $\Delta y=0.25$ and $Y=200$, on the last quenching run, as a
function of the blocks of sweeps. The vertical scale is relative to the mean
value $\bar{y}_{0-3}=18.02868690614052$, which corresponds to the zero of the
scale.

\item\label{fig07}
Binding energy $E_{0l}$, in $eV$, versus the dimensionless parameter
$\beta_{in}$, for $l=-1$ and $l=-3$.

\item\label{fig08}
Average radius $\bar{r}_{0l}$, in \AA, versus the dimensionless parameter
$\beta_{in}$, for $l=-1$ and $l=-3$.

\item\label{fig09}
Dissociation temperature $T_{d,0l}$, in $K$, versus the dimensionless parameter
$\beta_{in}$, for $l=-1$ and $l=-3$.

\item\label{fig10}
Resonant frequency $f_R$, in $MHz$, versus the dimensionless parameter
$\beta_{in}$, for the ($l=-3$)---($l=-5$), $n=0$ doublet.

\item\label{fig11}
Resonant frequency $f_R$, in $MHz$, versus the dissociation temperature
$T_{d,0l}$, in $K$, for $l=-3$.

\end{enumerate}

\newpage

\begin{center}
TABLES
\end{center}

\begin{enumerate}

\item\label{tbl1}
Approximate fits for the numerical data.

\end{enumerate}

\begin{center}
\begin{tabular}{c|l|l}
\hline\hline
Graph\rule{0em}{2.4ex}
& \multicolumn{2}{|c}{Fits} \\
\hline\hline
$E_{0l}\,\times\,\beta_{in}$\rule{0em}{2.4ex}
& $E_{0-1}\approx 2.25\times 10^4\;\beta_{in}^{-2.224}\;eV$
& $E_{0-3}\approx 9.55\times 10^4\;\beta_{in}^{-2.229}\;eV$ \\
\hline
$\bar{r}_{0l}\,\times\,\beta_{in}$\rule{0em}{2.4ex}
& $\bar{r}_{0-1}\approx 4.05\times 10^{-2}\;\beta_{in}^{1.113}$ \AA
& $\bar{r}_{0-3}\approx 2.86\times 10^{-2}\;\beta_{in}^{1.117}$ \AA \\
\hline
$T_{d,0l}\,\times\,\beta_{in}$\rule{0em}{2.4ex}
& $T_{d,0-1}\approx 2.61\times 10^8\;\beta_{in}^{-2.224}\;K$
& $T_{d,0-3}\approx 1.11\times 10^9\;\beta_{in}^{-2.229}\;K$ \\
\hline
$f_R\,\times\,\beta_{in}$\rule{0em}{2.4ex}
& \multicolumn{2}{|c}{$f_R\approx 1.29\times 10^{13}
\;\beta_{in}^{-3.465}\;MHz$} \\
\hline
$f_R\,\times\,T_{d,0-3}$\rule{0em}{2.4ex}
& \multicolumn{2}{|c}{$f_R\approx 1.13\times 10^{-1}
\;T_{d,0-3}^{1.554}(K)\;MHz$} \\
\hline\hline
\end{tabular}
\end{center}

\end{document}